\documentclass[reprint,amsmath,amssymb,aps,pra]{revtex4-2}
\usepackage{siunitx}
\usepackage{graphicx}
\usepackage{bm}
\usepackage{braket}
\usepackage{hyperref}
\usepackage{cleveref}
\usepackage{amsfonts}
\usepackage{amsthm}
\usepackage{mathtools}
\usepackage{microtype}
\usepackage{float}
\usepackage{comment}

\theoremstyle{remark}

\usepackage[dvipsnames]{xcolor}
\usepackage{colortbl}
\usepackage{stmaryrd}
\usepackage{tcolorbox}
\definecolor{CorrectionBlue}{RGB}{0,64,180}

\newcommand{\reviewfigure}[1]{%
  \IfFileExists{#1}{\includegraphics[width=0.45\textwidth]{#1}}{%
    \fbox{\parbox[c][2.2cm][c]{0.40\textwidth}{\centering
    \color{CorrectionBlue}Figure not supplied:\\ \texttt{#1}\\
    Original caption retained; plot not verified.}}}}
\begin{document}

\title{Orbital Angular Momentum Transfer and Beam-Quality Limits in Fork-Grating Electron Diffraction}

\author{P.K. Tsiopa}
\affiliation{School of Physics and Engineering,
ITMO University, St. Petersburg, Russia 197101}%

\author{S.S. Baturin}
\affiliation{School of Physics and Engineering,
ITMO University, St. Petersburg, Russia 197101}%

\date{\today}
\begin{abstract}
We analyze fork-grating generation of electron vortex beams beyond the ideal
single-mode approximation. The incident beam is described as a paraxial
Gaussian ensemble with finite transverse momentum spread and a first
electromagnetic Hartree phase accumulated before the grating. Within Fresnel
propagation we derive an analytical expression for each diffraction order
and determine the corresponding orbital-angular-momentum statistics. The
mean OAM of a selected order remains fixed by the topological rule,
$\langle L_z\rangle_m=mp\hbar$, while its variance is set by the residual
transverse momentum spread of the beam prepared at the grating. For the
Gaussian mixture, the transverse purity is
$\mathcal P_{\perp,m}=[1+2(\Delta L_{z,m}/\hbar)^2]^{-1}$. The observed
annular profile is governed by a separate propagation-dependent blurring
scale, so a beam can retain well-defined mean OAM while having low transverse
purity or a partially filled vortex minimum. Within the radial phase-screen
approximation, the Hartree self-field modifies the propagated profile while
preserving the full-plane OAM moments and transverse purity.
\end{abstract}

\maketitle

\section{Introduction}

Electron vortex beams provide access to electron states with controlled orbital angular momentum (OAM) and have motivated research in electron microscopy, scattering, and quantum-state preparation \cite{bliokh2017,lloyd2017}. Their extension to accelerator energies has stimulated studies of vortex-state capture and acceleration in electromagnetic fields \cite{baturin2022,meng2025}. Analytical descriptions of quantum transport through nonuniform magnetic optics also provide a framework for propagating structured electron states through magnetic lenses \cite{filina2026}. The state prepared for injection is characterized by its transverse coherence, concentration around the intended OAM value, and available current. Quantifying these properties is an essential step toward accelerator applications of vortex electrons.

Interest in preparing vortex electrons for accelerator applications is reflected in studies of RF photoinjectors \cite{dyatlov2025}, ultraviolet optical vortex generation with photoinjector lasers \cite{Dyatlov:25}, and proposed grating-based field-emission sources \cite{China}.

Holographic fork gratings offer an established method for generating electron vortex beams \cite{ushida2010,verbeeck2010,McMorran2011}. For centered, rotationally symmetric coherent illumination, a grating with signed fork charge $p$ introduces the azimuthal factor $e^{-imp\varphi}$ into diffraction order $m$. In the reference frame of that order, the ideal field carries OAM $mp\hbar$ in the convention used here. The quality of the generated state depends on the illumination prepared at the mask. Its available current is further constrained by diffraction efficiency, losses in the grating, and the apertures used to prepare the illumination and isolate the selected order \cite{harvey2014}. A quantitative description of grating generation must therefore account for the angular spread and transverse coherence of the transmitted beam.

The illumination reaching the grating can comprise an incoherent ensemble of modes with different transverse tilts. The role of partial coherence in electron vortex formation has been studied in Ref.~\cite{schattschneider2011}. Their distribution depends on the source and beam preparation, with mean transverse energy (MTE) and emittance providing relevant source and transport scales \cite{dowell2009,bazarov2009,karkare2017,karkare2020,vanOudheusden2010coherence,alberdi2022}. The tilts broaden the OAM distribution about a common order axis and reduce transverse-state purity. They also produce displaced diffraction patterns downstream, whose incoherent average can fill the vortex intensity minimum. At sufficiently high current density, Coulomb self-fields introduce additional changes to the incident beam \cite{luiten2004,vanoudheusden2010}. OAM statistics, transverse-state purity, and intensity visibility thus provide complementary measures of the state prepared for subsequent transport.

In this work, we extend the analytical Gaussian fork-diffraction framework of Ref.~\cite{janicijevic2008} to a paraxial electron ensemble with a Gaussian distribution of residual transverse tilts and a radial electromagnetic Hartree phase screen before the grating. We derive simple analytical formulas for the OAM dispersion and transverse-state purity of the Gaussian tilt mixture. The common radial Hartree phase modifies the propagated intensity profile and preserves the full-plane OAM moments and transverse-state purity. We show that the purity, defined as $\mathcal P_{\perp,m}=\operatorname{Tr}(\hat\rho_m^2)$, is determined by the absolute dispersion through $\mathcal P_{\perp,m}=[1+2(\Delta L_{z,m}/\hbar)^2]^{-1}$. A small relative dispersion can therefore coexist with low purity. In the regime $\hbar\ll\Delta L_{z,m}\ll|mp|\hbar$, the OAM fluctuations are small compared with the mean and span many quantum units, giving the OAM statistics a semiclassical character. The resulting formulas quantify both the relative OAM spread and the degree of statistical mixing of the transverse state.

\section{Source and paraxial mean-field model}
\label{sec:model}

We consider electrons emitted from a finite transverse source, accelerated to a dominant longitudinal momentum, and incident on a fork grating placed at $z=\zeta$. The electric and magnetic fields retained below are not externally applied lens fields. They are the ensemble-averaged fields generated by all other electrons in the incident beam, and they act only in the source-to-grating interval $0<z<\zeta$.

We use the phase convention $\exp[i(Et-\bm p\cdot\bm r)/\hbar]$,
with $\hat E=-i\hbar\partial_t$ and $\hat{\bm p}=+i\hbar\nabla$.
The Dirac matrices $\bm\alpha,\beta$, and the Pauli matrices
$\bm\sigma$ below are the elementwise complex conjugates of those
in the standard Dirac representation. This is a change of representation,
not charge conjugation: the electron charge remains $-e$, with $e>0$.
In the quasistatic electromagnetic Hartree approximation, a representative transverse-momentum component obeys
\begin{align}
-i\hbar\partial_t\psi_{\bm k_\perp}
&=\left[c\bm\alpha\cdot\hat{\bm\pi}_H
+\beta m_ec^2-e\Phi_H\right]\psi_{\bm k_\perp},
\notag\\
\hat{\bm\pi}_H&=+i\hbar\nabla+e\bm A_H.
\label{eq:dirac_mean_field}
\end{align}
In the Coulomb gauge, neglecting retardation and radiation, the ensemble-averaged potentials satisfy
\begin{align}
\nabla^2\Phi_H&=\frac{e}{\varepsilon_0}n,
&\nabla^2\bm A_H&=-\mu_0\bm j_H,
\label{eq:maxwell_hartree_fields}
\end{align}
\begin{align}
n&=\sum_s\int d^2k_\perp\,P(\bm k_\perp)
\psi_{\bm k_\perp s}^\dagger\psi_{\bm k_\perp s},
\notag\\
\bm j_H&=-e\sum_s\int d^2k_\perp\,P(\bm k_\perp)
\psi_{\bm k_\perp s}^\dagger c\bm\alpha\psi_{\bm k_\perp s}.
\label{eq:maxwell_hartree_sources}
\end{align}

The spinors in these source equations are density-normalized, with spin
population weights absorbed into their amplitudes. The Gaussian fields below
specify relative mode shapes. Their arbitrary overall amplitude is not an
independent determination of the physical density $n_0$.

Thus each electron propagates in the average field of the remaining beam. The distribution $P(\bm k_\perp)$ labels the incident transverse momenta, while both mean potentials are generated by the total ensemble density and current. Equations~\eqref{eq:maxwell_hartree_fields} and \eqref{eq:maxwell_hartree_sources} define the first approximation used below: exchange and correlations beyond the direct Hartree field are omitted, and the field is treated quasistatically, without retardation or radiation.

We reduce Eq.~\eqref{eq:dirac_mean_field} directly, without a nonrelativistic Foldy-Wouthuysen expansion. Exact elimination of the lower spinor component produces, in addition to the orbital kinetic term, Zeeman, electric spin-orbit, and spin-independent electric-gradient corrections. Their explicit forms are given in Appendix~\ref{app:relativistic_reduction}.

The smallness of the spin and electric-gradient terms depends on energy,
beam scale, and transverse momentum, as quantified in
Eq.~\eqref{eq:operator_estimates_app}. At $U=\SI{10}{kV}$ and
$w_0=10^{-7}$-$10^{-6}\,\mathrm{m}$, the individual Zeeman-to-electric
ratios are approximately $4\times10^{-7}$-$4\times10^{-8}$, and the
spin-orbit ratios approximately $2\times10^{-7}$-$2\times10^{-8}$.
The scalar-gradient ratio is approximately
$4\times10^{-12}$-$4\times10^{-14}$ only for
$p_\perp\sim\hbar/w_0$; a large component tilt requires the actual
$p_\perp$ in that estimate. These are ratios to the electric term,
not universal bounds relative to the magnetically reduced net self-field.
Subject to these smallness conditions, we omit the corrections and retain
both orbital mean potentials. The upper component then obeys

\begin{equation}
\hat{\bm\pi}_H^2\varphi
=\frac{(E+e\Phi_H)^2-m_e^2c^4}{c^2}\varphi.
\label{eq:scalar_relativistic_main}
\end{equation}
This is a spinless relativistic Helmholtz equation, not a nonrelativistic limit. For an unpolarized source, contributions linear in spin average to zero; spin incoherence alone need not imply zero polarization. Neglect of the spin operators still requires the preceding smallness estimates.

For the stationary, nearly cylindrical beam considered here,
\begin{align}
\bm A_H=A_{H,z}(r)\bm e_z,
\qquad&
\bm B_H=-\partial_rA_{H,z}\bm e_\varphi,
\notag\\
&j_{H,z}\simeq-en v_z.
\label{eq:self_field_geometry}
\end{align}
Thus the self-magnetic field is azimuthal. Transverse currents and longitudinal variations of the mean field are neglected at paraxial order.

Let
\begin{equation}
E^2=m_e^2c^4+\hbar^2c^2k^2,
~~
p_z=\hbar k,
~~
v_z=\frac{p_zc^2}{E},
\label{eq:relativistic_kinematics}
\end{equation}
and set $\varphi_s=u(\bm r_\perp,z)e^{-ikz}\eta_s$, where $\eta_s$ is a constant two-component spinor with $\eta_s^\dagger\eta_{s'}=\delta_{ss'}$. Here $u$ is the physical envelope; the spin index is suppressed below. With the stated momentum operator, $k>0$ corresponds to $p_z=\hbar k>0$. We retain the exact $E$-$k$ relation and assume weak mean fields and a narrow angular spectrum,
\begin{equation}
\frac{e|\Phi_H|}{E}\ll1,
~~
\frac{e|A_{H,z}|}{p_z}\ll1,
~~
\frac{|\bm k_\perp|}{k}\ll1.
\label{eq:paraxial_condition_main}
\end{equation}

The slowly varying longitudinal envelope also requires
\begin{equation}
\frac{e|\Phi_H-v_z A_{H,z}|}{p_zv_z}\ll1.
\label{eq:longitudinal_field_condition}
\end{equation}
Potential offsets are fixed consistently with the reference carrier.
Paraxiality must hold for the coherent Gaussian divergence and the
selected grating kick as well:
$1/(kw_0)\ll1$ and $|2\pi m/(kD)|\ll1$, in addition to the residual
angular spread. The Gaussian momentum distribution is understood to have
negligible weight outside this range.

Together with the slowly varying envelope condition, these assumptions remove terms quadratic in the potentials, the second longitudinal derivative, and mixed weak-field-paraxial corrections. Their explicit form is given in Appendix~\ref{app:relativistic_reduction}. At the retained order,
\begin{equation}
\partial_z u=-\frac{i}{2k}\nabla_\perp^2u
-\frac{ie}{\hbar v_z}\left(\Phi_H-v_zA_{H,z}\right)u.
\label{eq:paraxial_sp}
\end{equation}
The free term has the usual Fresnel form, while relativity remains in $k$, $v_z$, and the combination $\Phi_H-v_zA_{H,z}$. The free Foldy-Wouthuysen $\hat{\bm p}^{\,4}$ correction is already contained in the exact dispersion; the first omitted free term is instead a nonparaxial transverse correction, as shown in Appendix~\ref{app:relativistic_reduction}.

A representative photoemission-based scale is an accelerating voltage of order $U\sim\SI{10}{kV}$, a laser spot diameter of order $d_L\sim\SI{200}{nm}$, and currents in the nA-$\mu$A range. Relativistically,
\begin{align}
p_z&=\frac{1}{c}\sqrt{eU(eU+2m_ec^2)}, \nonumber\\
v_z&=c\frac{\sqrt{eU(eU+2m_ec^2)}}{m_ec^2+eU},
\label{eq:p_v_estimate}
\end{align}
which gives $p_z\approx\SI{5.4e-23}{kg.m/s}$, $v_z\approx\SI{5.8e7}{m/s}$, and $k\approx\SI{5.1e11}{m^{-1}}$ for $U=\SI{10}{kV}$. For a beam radius $r\sim\SI{100}{nm}$, current continuity $I=nev_z\pi r^2$ gives densities in the approximate range $n\sim10^{15}$-$10^{18}\,\mathrm{m}^{-3}$. These numbers are used only as a scale-setting example; the subsequent formulae are written so that $w_0$, $\sigma_k$, $D$, $\zeta$, and $z-\zeta$ can be replaced by the parameters of a particular apparatus.

For the grating itself we keep $D$ explicit, but the values used in electron-vortex experiments are commonly in the sub-\SI{100}{nm} to micron range depending on fabrication and beam energy. The transverse grating momentum $2\pi/D$ should therefore be treated on the same footing as the source transverse momentum $\sigma_k$: even if the incoming electrons were perfectly collimated, the selected off-axis diffraction order is displaced by the grating. This displacement is one of the reasons why OAM dispersion and screen-profile blurring are not identical notions.

The transverse momentum width is more naturally connected with the source MTE than with an equilibrium Fermi momentum. A formal density-based Fermi wave number, $k_F=(3\pi^2n)^{1/3}$, gives only $10^5$-$10^6\,\mathrm{m}^{-1}$ for the above densities. By contrast, the MTE estimate

\begin{align}
\mathrm{MTE}_{\rm mix}
&=\frac{\hbar^2\langle k_x^2+k_y^2\rangle}{2m_e}
=\frac{\hbar^2\sigma_k^2}{2m_e},\notag\\
\sigma_k&\simeq\frac{\sqrt{2m_e\,\mathrm{MTE}}}{\hbar}.
\label{eq:mte_sigmak}
\end{align}
Here the estimate identifies the measured MTE with the energy of the
random component tilts. Each Cartesian rms wave number is
$\sigma_k/\sqrt{2}$. If the intrinsic Gaussian momentum width is retained
at its waist, the full mode-mixture energy instead is
\begin{equation}
\mathrm{MTE}_{\rm total}
=\frac{\hbar^2}{2m_e}\left(\sigma_k^2+\frac{2}{w_0^2}\right).
\label{eq:mte_total}
\end{equation}
Deterministic beam steering is excluded from these spreads. The tilt-dominated estimate corresponds to $\sigma_k\sim0.81\times10^9$-$3.62\times10^9\,\mathrm{m}^{-1}$ for MTE values in the range $0.025$-$0.5\,\mathrm{eV}$ \cite{dowell2009,bazarov2009,karkare2017,karkare2020}. The smaller values used below, such as $\sigma_k\sim10^7\,\mathrm{m}^{-1}$, should therefore be understood as the effective angular width of the collimated fraction reaching the fork grating, not as the full source MTE.

We finally evaluate the Hartree potentials once on the noninteracting beam and retain their action as an eikonal phase screen before the grating. The precise content of this first-Hartree approximation is stated in Appendix~\ref{app:relativistic_reduction}. The incident field is then
\begin{align}
&u_{\bm k_\perp}^{(1)}(x,y,z)\simeq u_{\bm k_\perp}^{(0)}(x,y,z)
\nonumber \\
&\times\exp\left[-\frac{ie}{\hbar v_z}
\int_0^z\left(\Phi_H^{(0)}-v_zA_{H,z}^{(0)}\right)dz'\right].
\label{eq:eikonal_hartree_main}
\end{align}
For a uniform cylindrical beam of laboratory density $n_0$, the quasistatic Maxwell equations give
\begin{align}
\Phi_H^{(0)}(r)&=\Phi_H^{(0)}(0)+\frac{en_0}{4\varepsilon_0}r^2,
\notag\\
A_{H,z}^{(0)}(r)&=A_{H,z}^{(0)}(0)+\frac{\mu_0en_0v_z}{4}r^2.
\label{eq:uniform_hartree_potentials}
\end{align}

These are interior potentials of a uniform-density cylinder, an additional
approximation to the actual Gaussian ensemble. For example, a cylindrical
Gaussian density $n(r)=n_0e^{-2r^2/w_g^2}$ gives
\begin{equation}
\partial_r\Phi_H
=\frac{en_0w_g^2}{4\varepsilon_0r}
\left(1-e^{-2r^2/w_g^2}\right),
\label{eq:gaussian_self_field}
\end{equation}
whose near-axis potential is
$\Phi_H(0)+en_0r^2/(4\varepsilon_0)
-en_0r^4/(8\varepsilon_0w_g^2)+\cdots$.
Consequently, the quadratic screen used below is a uniform-density or
near-axis model. 

Hence
\begin{equation}
\Phi_H^{(0)}-v_zA_{H,z}^{(0)}
=\text{const}+\frac{en_0}{4\varepsilon_0\gamma^2}r^2,
~~
\gamma=\frac{E}{m_ec^2}.
\label{eq:effective_self_potential}
\end{equation}
The factor $1/\gamma^2=1-v_z^2/c^2$ expresses the cancellation between the electric space-charge force and the magnetic force of the beam current. This simple factor assumes a long quasistationary beam with approximately $z$-independent laboratory density and current. For a short or longitudinally structured bunch, the two potentials in Eq.~\eqref{eq:eikonal_hartree_main} must instead be evaluated separately. Taking the interaction length before the grating to be $\zeta$ gives
\begin{align}
\chi_{\rm em}(r,\zeta)&=\frac{e}{\hbar v_z}
\int_0^\zeta\left(\Phi_H^{(0)}-v_zA_{H,z}^{(0)}\right)dz'
\notag\\
&=\chi_0+\tilde\chi_{\rm em}r^2,
\notag\\
\tilde\chi_{\rm em}
&=\frac{e^2n_0\zeta}{4\varepsilon_0\hbar v_z\gamma^2}
=\frac{e^2m_en_0\zeta}{4\varepsilon_0\hbar^2k\gamma}.
\label{eq:hartree_phase}
\end{align}
The global phase $\chi_0$ is dropped. For $\gamma\to1$, Eq.~\eqref{eq:hartree_phase} reduces to the usual electrostatic Schr\"odinger-Poisson result.
In the chosen convention the accumulated phase is
$-\tilde\chi_{\rm em}r^2$. Since $\hat{\bm p}=+i\hbar\nabla$,
it gives the outward kick $\Delta p_r=2\hbar\tilde\chi_{\rm em}r$:
the electric self-field is defocusing and the magnetic field reduces it.

The following Gaussian and diffracted fields include the carrier
$e^{-ikz}$, whereas $u$ in Eq.~\eqref{eq:paraxial_sp} is its envelope.
The untilted reference mode is represented by a Gaussian paraxial field
\begin{align}
U^{(i)}(r,z)&=\frac{q(0)}{q(z)}
\exp\left[-ik\left(z+\frac{r^2}{2q(z)}\right)\right],
\notag\\
q(z)&=z+iz_R,
\label{eq:gaussian_mode}
\end{align}
where $z_R=kw_0^2/2$. The incident ensemble is described by a transverse momentum distribution
\begin{equation}
P(\bm k_\perp)=\frac{1}{\pi\sigma_k^2}
\exp\left[-\frac{k_x^2+k_y^2}{\sigma_k^2}\right],
\label{eq:momentum_distribution}
\end{equation}
with $\langle k_x^2\rangle=\langle k_y^2\rangle=\sigma_k^2/2$.

The statistical boundary condition is imposed at the grating. Namely, the
components share one centered Gaussian amplitude and curvature and differ
by their linear phases. Here $w_0$ and $\zeta$ parameterize a reference
Gaussian waist, and $\sigma_k$ is the residual tilt width of the prepared
mixture at the mask. This is not automatically the state obtained by
ballistic propagation of tilted packets from a common source waist.
Indeed, free propagation from that waist gives
\begin{align}
&u_{\bm k_\perp}^{(0)}(\bm r_\perp,\zeta)
 = \nonumber \\
 &e^{-i\bm k_\perp\cdot\bm r_\perp
       +i\zeta k_\perp^2/(2k)}
 u_0^{(0)}\!\left(\bm r_\perp-
             \frac{\zeta\bm k_\perp}{k},\zeta\right),
\label{eq:ballistic_tilt}
\end{align}
with $u_0^{(0)}$ the untilted envelope. The common-amplitude approximation
therefore requires appropriate beam preparation or negligible walkoff,
$\zeta\sigma_k/k\ll w_g$, where
$w_g=w_0\sqrt{1+\zeta^2/z_R^2}$. The same issue applies to deterministic
steering, i.e. the mean beam must illuminate the fork center.

Thus the field incident on the grating is
\begin{align}
u_{\rm in}(r,\varphi,\zeta)=&\,\frac{q(0)}{q(\zeta)}
\exp\left[-ik\left(\zeta+\frac{r^2}{2q(\zeta)}\right)\right]
\notag\\
&\times\exp\left(-i\bm k_\perp\!\cdot\!\bm r_\perp-i\tilde\chi_{\rm em} r^2\right).
\label{eq:incident_field}
\end{align}
Equation~\eqref{eq:incident_field} is the field immediately before the grating. The self-field is not propagated through the diffraction region: after the phase has accumulated over $0<z<\zeta$, the mask is applied and each selected order is treated as freely propagating.
\section{Fork-grating diffraction}
\label{sec:diffraction}
The fork transmission function is expanded into diffraction orders,
\begin{equation}
T(r,\varphi)=\sum_{m=-\infty}^{\infty}t_m
\exp\left[-im\left(\frac{2\pi}{D}r\cos\varphi+p\varphi\right)\right],
\label{eq:fork_transmission}
\end{equation}
where $D$ is the grating period, $p$ is the signed fork charge, and $t_m$ is the complex amplitude of the $m$th order. We choose the sign of $p$ so that $\hat L_z=+i\hbar\partial_\varphi$ assigns OAM $mp\hbar$ to $e^{-imp\varphi}$; thus $p$ is opposite in sign to the parameter in the transmission function of Ref.~\cite{janicijevic2008}. The field after the mask is propagated by the Fresnel kernel from $\zeta$ to $z$. Combining the Gaussian curvature, free propagation, and Hartree phase gives
\begin{equation}
\tilde\gamma=\frac{1}{q(\zeta)}+\frac{1}{z-\zeta}+\frac{2\tilde\chi_{\rm em}}{k},
~~
Q(z)=\frac{(z-\zeta)^2\tilde\gamma}{2}.
\label{eq:gamma_q}
\end{equation}

\begin{figure}[t]
    \centering
    \reviewfigure{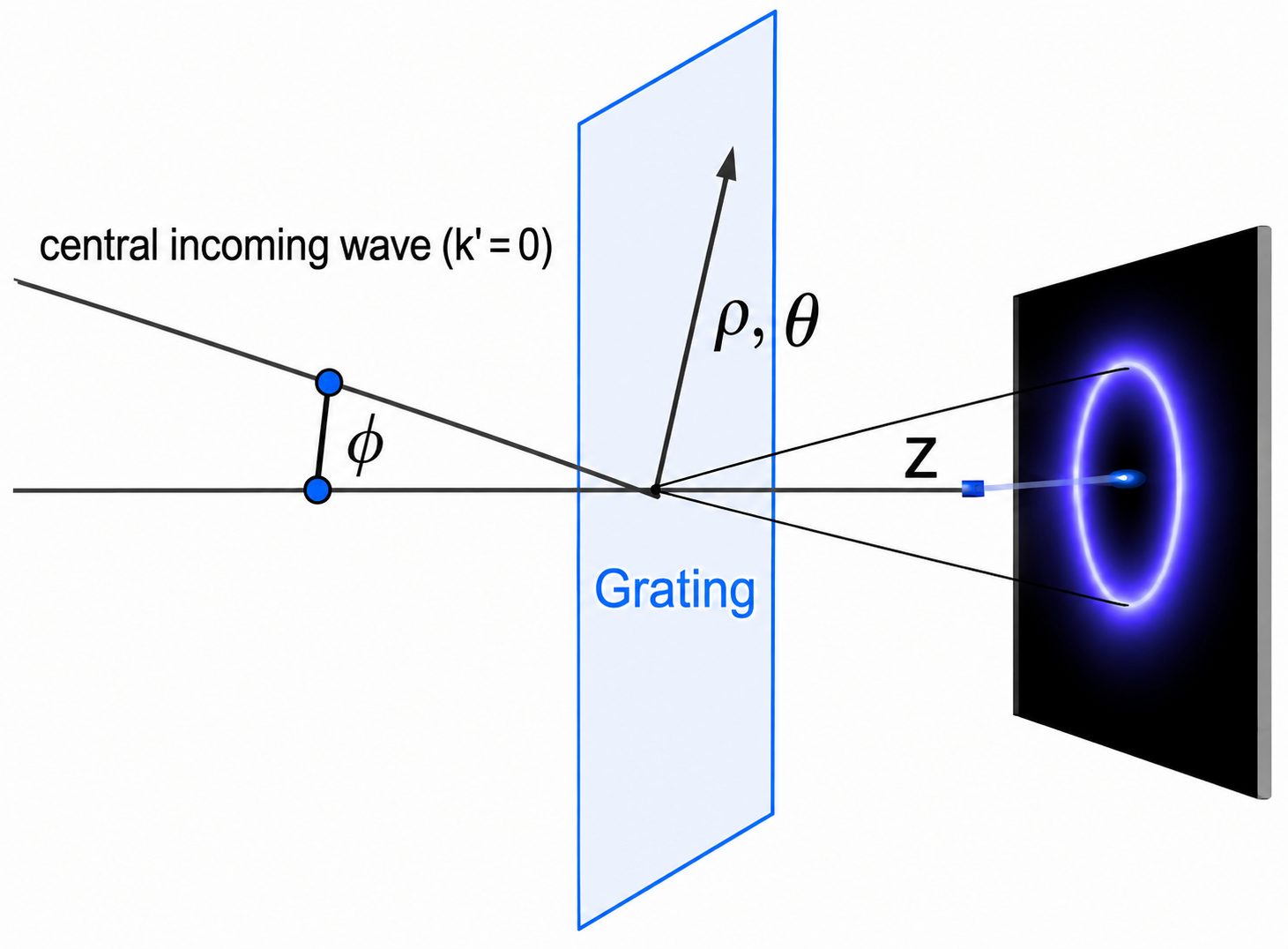}
    \caption{Geometry of fork-grating preparation with compensation of the
selected order's deterministic transverse momentum. The fixed incident
angle $\phi$ is chosen so that $\bm K_m=0$ in
Eq.~\eqref{eq:net_order_tilt}; the mean component of that order then
propagates along the grating normal $z$. The coordinates $(\rho,\theta)$
are measured about this common axis. The notation $k'=0$ in the schematic
denotes zero residual transverse wave vector, $\bm k_\perp=0$.}
    \label{fig:mean}
\end{figure}

For a chosen order $m_0$, define the deterministic incident tilt
$\bm k_0=k_{x,0}\bm e_x$ and the net transverse wave vector of order $m$,
\begin{equation}
\bm K_m=\bm k_0+\frac{2\pi m}{D}\bm e_x.
\label{eq:net_order_tilt}
\end{equation}
The incident field in Eq.~\eqref{eq:incident_field} must be multiplied by
$e^{-i\bm k_0\cdot\bm r_\perp}$. Choosing
$k_{x,0}=-2\pi m_0/D$ centers that selected order, for which
$\bm K_{m_0}=0$ and
$\sin\phi\simeq-2\pi m_0/(kD)$ as in Fig.~\ref{fig:mean}.
The incident tilt is fixed once for the whole beam; it cannot depend on
the summation index in the grating transmission.
In the centered-order formulas below, $m$ denotes this selected order
and $\bm k_\perp$ is its residual incident transverse wave vector.

For the mean residual component, the selected-order center then coincides
with the laboratory axis. For a component with residual wave vector
$\bm k_\perp$, introduce local coordinates $(\rho_m,\theta_m)$:

\begin{equation}
\rho_m\cos\theta_m+\alpha_m=\rho\cos\theta,
~~
\rho_m\sin\theta_m+\beta_m=\rho\sin\theta.
\label{eq:shifted_coords}
\end{equation}
For the compensated selected order,
\begin{equation}
\alpha_m=(z-\zeta)\frac{k_x}{k},
\qquad
\beta_m=(z-\zeta)\frac{k_y}{k},
\label{eq:order_shift}
\end{equation}
For normal incidence, $\bm k_0=0$, and the laboratory displacement is instead
\begin{equation}
\alpha'_m=(z-\zeta)\left(\frac{2\pi m}{kD}+\frac{k_x}{k}\right),
~~
\beta'_m=(z-\zeta)\frac{k_y}{k},
\label{eq:order_shift_perp}
\end{equation}
The mean component of an off-axis order is therefore displaced even when $\bm k_\perp=0$. For arbitrary fixed incident tilt, use the total shift $(z-\zeta)(\bm K_m+\bm k_\perp)/k$ in Eq.~\eqref{eq:shifted_coords} and in the order-resolved field.

\begin{widetext}
After the grating the field is
\begin{align}
u(\rho,\theta,z)=&\,\frac{ik e^{-ik(z-\zeta)}}{2\pi(z-\zeta)}
\int d^2r\,u_{\rm in}(r,\varphi,\zeta)T(r,\varphi)
\exp\left[-\frac{ik}{2(z-\zeta)}|\bm r_\perp-\bm\rho|^2\right].
\label{eq:fresnel_app}
\end{align}

The angular part of the $m$th term is reduced by shifting the linear phase into the local variables \eqref{eq:shifted_coords}. The standard identity
\begin{align}
\int_0^{2\pi}e^{ia r\cos(\varphi-\theta_m)}e^{-imp\varphi}\,d\varphi
=2\pi i^{|mp|}J_{|mp|}(ar)e^{-imp\theta_m}
\label{eq:angular_integral_app}
\end{align}
leaves the radial integral 
\begin{equation}
Y_m(\rho_m)=\int_0^\infty J_{|mp|}\left(\frac{k\rho_m r}{z-\zeta}\right)
\exp\left[-\frac{ik\tilde\gamma r^2}{2}\right]r\,dr.
\label{eq:radial_integral_app}
\end{equation}
Evaluation of Eq.~\eqref{eq:radial_integral_app} first gives a Kummer-function form and then, after the standard transformation used in optical vortex diffraction, the Bessel representation. Therefore the propagated $m$th order is
\begin{align}
u_m(\rho_m,\theta_m,z)=&\,t_m\frac{q(0)}{q(\zeta)}
\frac{\sqrt{\pi k}(z-\zeta)}{8Q^{3/2}(z)}i^{|mp|/2}
\exp\left\{-ik\left[z+\frac{\rho^2}{2(z-\zeta)}\right]\right\}
e^{-imp\theta_m}\rho_m
\exp\left[\frac{ik\rho_m^2}{8Q(z)}\right]
F\!\left(\frac{k\rho_m^2}{8Q(z)}\right),
\label{eq:um_final}
\end{align}
where
\begin{equation}
F(X)=J_{(|mp|-1)/2}(X)+iJ_{(|mp|+1)/2}(X).
\label{eq:F_def}
\end{equation}
Equation~\eqref{eq:um_final} is the order-resolved electron analogue of the optical fork-diffraction result. The factor $\exp(-imp\theta_m)$ contains the vortex phase in the stated convention, while Eqs.~\eqref{eq:order_shift} and \eqref{eq:shifted_coords} show how a real transverse momentum component is displaced on the observation plane. Fractional powers follow the branch obtained from $A=ik\tilde\gamma/2$ with $\operatorname{Re}A>0$. The expression follows from an intermediate confluent-hypergeometric form of the radial Fresnel integral. The full calculations following ~\cite{janicijevic2008} are presented in Appendix \ref{app:fresnel}

\end{widetext}

\begin{figure}[b]
    \centering
    \reviewfigure{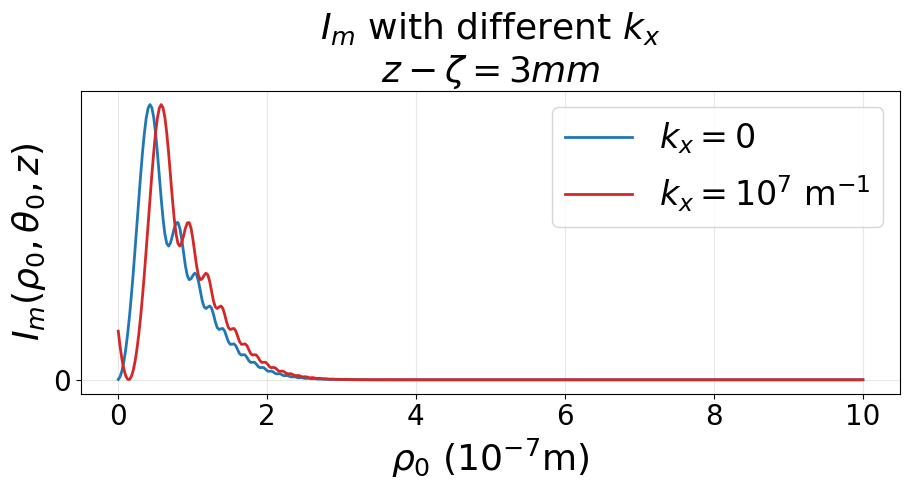}
    \caption{Displacement of single-component diffraction profiles for $m=1$ due to
residual transverse tilt. The blue and red curves correspond to
$k_x=0$ and $k_x=10^7\,\mathrm{m}^{-1}$, respectively, with $k_y=0$, at
$z-\zeta=3\,\mathrm{mm}$ in the low-density regime described in the text.
The profiles of Eq.~\eqref{eq:single_profile} are shown along the positive
$x$ axis of the compensated selected order; the plotting labels
$(\rho_0,\theta_0)$ denote the common coordinates $(\rho,\theta)$.
The tilted component is displaced by
$\alpha_m=(z-\zeta)k_x/k$ [Eq.~\eqref{eq:order_shift}], and its vortex
zero remains at its own displaced center.}
    \label{fig:shift}
\end{figure}

\begin{figure}[b]
    \centering
    \reviewfigure{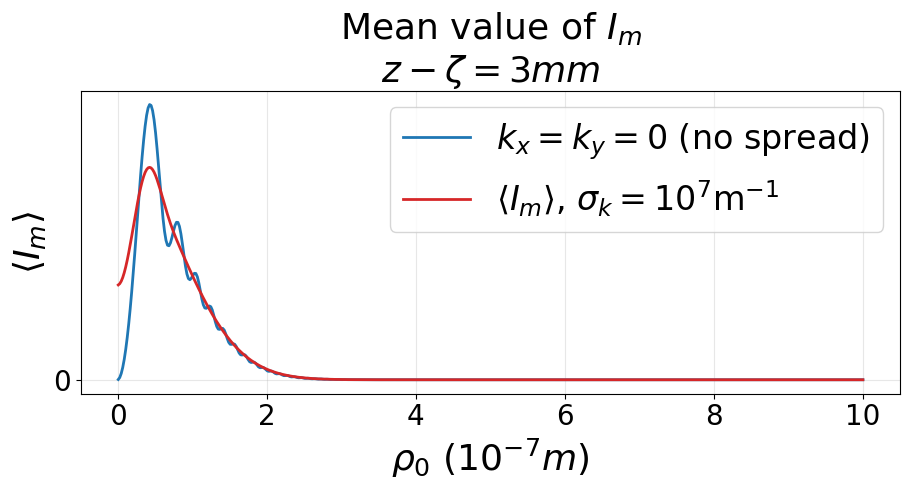}
    \caption{Effect of incoherent transverse-momentum averaging on the
observed diffraction profile for $m=1$ at $z-\zeta=3\,\mathrm{mm}$, with negligible
Hartree correction. The blue curve is the centered component
$k_x=k_y=0$; the red curve is the ensemble profile of
Eq.~\eqref{eq:ensemble_profile} for the Gaussian distribution in
Eq.~\eqref{eq:momentum_distribution}, with
$\sigma_k=10^7\,\mathrm{m}^{-1}$. Both curves are line profiles along the
positive $x$ axis, with $\rho_0=\rho$ and $\theta_0=\theta=0$.
Averaging over the displaced component centers fills the intensity minimum
at the mean beam center and smooths the radial oscillations.}
    \label{fig:profile}
\end{figure}

\section{Radial scale and observed intensity}
\label{sec:profile}

For a single component with nonzero charge $mp\ne0$, the selected order has a vortex zero at its own center $\rho_m=0$ for $z>\zeta$. The zero-charge case has no vortex zero and is obtained by the regular limit of the Bessel formula. The observed profile of an ensemble is different because each component has a different center. The normalized single-component profile is
\begin{equation}
I_m(\bm r_\perp,z|\bm k_\perp)=
\frac{|u_m(\rho_m,\theta_m,z;\bm k_\perp)|^2}
{\int d^2r_\perp |u_m(\rho_m,\theta_m,z;\bm k_\perp)|^2},
\label{eq:single_profile}
\end{equation}
However in the coordinates of the center of the beam the profile for non-central electron becomes shifted. Let us illustrate that effect on the Fig.~\ref{fig:shift}. At first we took the concentration to be low enough for the interaction corrections are not visible ($n=10^{15}\,\mathrm{m}^{-3}$).

The measured incoherent profile is illustrated on the Fig.~\ref{fig:profile}
\begin{equation}
I_m(\bm r_\perp,z)=\int d^2k_\perp\,P(\bm k_\perp)
I_m(\bm r_\perp,z|\bm k_\perp).
\label{eq:ensemble_profile}
\end{equation}

\begin{figure}[t]
    \centering
    \reviewfigure{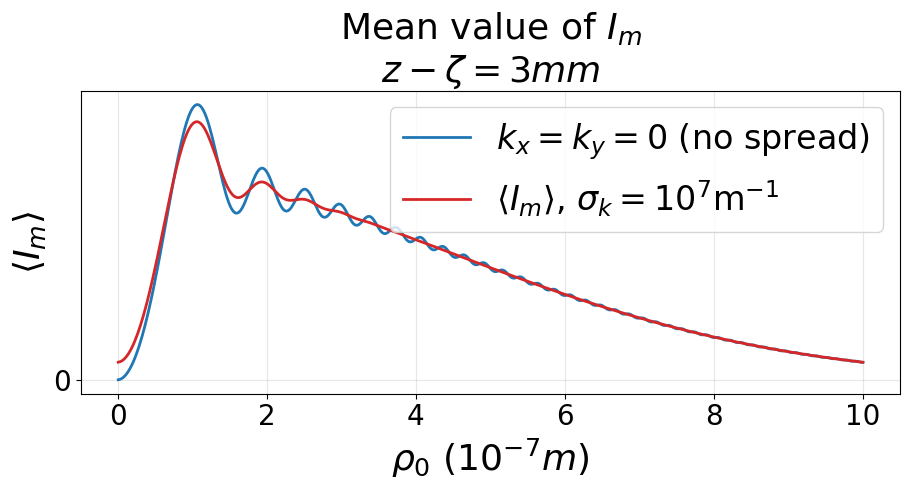}
    \caption{Diffraction profiles including the common radial electromagnetic
Hartree phase for $m=1$ of Eq.~\eqref{eq:hartree_phase}, at
$z-\zeta=3\,\mathrm{mm}$. The blue curve represents the centered component
$k_x=k_y=0$; the red curve represents the incoherent Gaussian ensemble
with $\sigma_k=10^7\,\mathrm{m}^{-1}$. The curves are plotted along the
positive $x$ axis, using $\rho_0=\rho$ and $\theta_0=\theta=0$.
The broader radial structure compared with Fig.~\ref{fig:profile}
illustrates the defocusing action of the Hartree phase. The scale
$\rho_{\rm c}$ in Eq.~\eqref{eq:rho_cross_general} characterizes the
exponential branch of the solution. This common radial phase preserves
the full-plane OAM moments and transverse-state purity. The concentration for this figure is taken as $n=10^{18}m^{-3}$ and $\tilde\chi_{\rm em}
\approx1.5\times10^{14}$.}
    \label{fig:inter}
\end{figure}

For nonzero Gaussian tilt width and $z>\zeta$, the convolution makes the ensemble intensity at the mean center strictly positive. Filling becomes substantial when the center spread is comparable with the actual ring radius.
A useful radial scale follows from the first order, $m=p=1$. For the first vortex order, the Hartree-modified exponential-envelope scale found in Appendix \ref{app:asymptotic_profile} is
\begin{equation}
\rho_{\rm c}=\sqrt{
\frac{\left|q(z)+\dfrac{2\tilde\chi_{\rm em}}{k}
(z-\zeta)q(\zeta)\right|^2}{4kz_R}
}.
\label{eq:rho_cross_general}
\end{equation}
Explicitly,
\begin{equation}
\rho_{\rm c}=\sqrt{
\frac{\left[z+\dfrac{2\tilde\chi_{\rm em}}{k}
\zeta(z-\zeta)\right]^2
+z_R^2\left[1+\dfrac{2\tilde\chi_{\rm em}}{k}
(z-\zeta)\right]^2}{4kz_R}
}.
\label{eq:rho_cross}
\end{equation}
For $\tilde\chi_{\rm em}=0$, this reduces to
$\rho_{\rm c}=\sqrt{(z^2+z_R^2)/(4kz_R)}$.
This scale characterizes the exponential branch, not a full intensity envelope or an exact ring radius. For $z>\zeta\ge0$ and $\tilde\chi_{\rm em}\ge0$, it increases with the defocusing screen strength. It does not by itself establish a numerical visibility threshold (Fig.~\ref{fig:inter}).

\section{Mean OAM and dispersion}
\label{sec:oam}
OAM is reference-frame dependent. For the compensated selected order,
$\bm K_m=0$, the order frame and the laboratory frame coincide. More
generally, put $L=z-\zeta$ and remove both the deterministic displacement
and momentum by defining
\begin{align}
\bm r'&=\bm r_\perp-\frac{L\bm K_m}{k},\notag\\
&u_m^{\rm com}(\bm r',z)
=\notag\\&e^{i\bm K_m\cdot\bm r'+iLK_m^2/(2k)}
 u_m^{\rm lab}\!\left(\bm r'+\frac{L\bm K_m}{k},z\right),
\label{eq:comoving_order_field}\\
\hat L_{z,m}^{\rm com}
&=\left[\left(\bm r_\perp-\frac{L\bm K_m}{k}\right)
\times\left(\hat{\bm p}_\perp-\hbar\bm K_m\right)\right]_z.
\label{eq:comoving_order_operator}
\end{align}
A translation of coordinates alone does not perform this momentum
subtraction. The following moments refer to this common order frame,
not a separately boosted frame for each random component. In that frame,
the residual component coordinates are shifted and the angular generator is

\begin{equation}
\frac{-i}{\hbar}\hat L_z\equiv D_m=N_-\partial_{\rho_m}
+\left(1+\frac{N_+}{\rho_m}\right)\partial_{\theta_m},
\label{eq:Dm}
\end{equation}
where
\begin{align}
N_+&=\alpha_m\cos\theta_m+\beta_m\sin\theta_m,
\notag\\
N_-&=\alpha_m\sin\theta_m-\beta_m\cos\theta_m.
\label{eq:Npm}
\end{align}
The terms that could shift the mean OAM are proportional to angular averages of sine and cosine harmonics and vanish. Hence
\begin{equation}
\langle L_z\rangle_m(\bm k_\perp)=mp\hbar,
\qquad
\langle L_z\rangle_m=mp\hbar.
\label{eq:mean_oam}
\end{equation}
This is the precise sense in which the topological transfer rule survives transverse spread and the first Hartree correction.

The second moment is not protected. It is defined as
\begin{equation}
\langle L_z^2\rangle_m(\bm k_\perp)=
\frac{\int d^2r_\perp\,u_m^*(-\hbar^2D_m^2)u_m}
{\int d^2r_\perp |u_m|^2},
\label{eq:lz2_def}
\end{equation}
or, equivalently when boundary terms are negligible,
\begin{equation}
\langle L_z^2\rangle_m(\bm k_\perp)=
\hbar^2\frac{\int d^2r_\perp |D_m u_m|^2}
{\int d^2r_\perp |u_m|^2}.
\label{eq:lz2_hermitian}
\end{equation}

The full calculations of the ~\eqref{eq:lz2_hermitian} are collected in Appendix \ref{app:oam_operator}. The final result is 
\begin{equation}
\frac{\Delta L_{z,m}^2}{\hbar^2}
=\frac{\sigma_k^2(\zeta^2+z_R^2)}{2kz_R}.
\label{eq:dispersion_final}
\end{equation}
For comparison, about the fixed laboratory axis through the fork center,
the same centered grating-plane mixture has
\begin{equation}
\frac{\Delta L_{z,m,\rm lab}^2}{\hbar^2}
=\frac{(\sigma_k^2+K_m^2)(\zeta^2+z_R^2)}{2kz_R}.
\label{eq:dispersion_lab}
\end{equation}
For normal incidence $K_m^2=(2\pi m/D)^2$. The deterministic momentum
therefore adds in quadrature to the residual two-dimensional rms wave
number 
In the
translated and momentum-boosted order frame of
Eqs.~\eqref{eq:comoving_order_field}-\eqref{eq:comoving_order_operator},
Eq.~\eqref{eq:dispersion_final} remains unchanged. These statements
assume full-plane integration and no selection that clips an order.

An illustrative parameter set is
\begin{align}
w_0&=10^{-7}\,\mathrm{m},
&L&=3\times10^{-3}\,\mathrm{m},\notag\\
D&=10^{-6}\,\mathrm{m},
&\sigma_k&=10^7\,\mathrm{m}^{-1},\notag\\
k&=2\times10^{12}\,\mathrm{m}^{-1},
&z_R&=10^{-2}\,\mathrm{m}.
\label{eq:params}
\end{align}
This wave number corresponds to a kinetic energy of approximately
$\SI{134.7}{keV}$, distinct from the earlier $\SI{10}{kV}$ example.
For the grating distance $\zeta>10$ mm these parameters imply
\begin{equation}
\sqrt{\frac{\Delta L_{z,m}^2}{\hbar^2}}
=0.5\sqrt{1+(\zeta/\SI{10}{mm})^2}\ge \frac{1}{\sqrt{2}},
\label{eq:example_oam}
\end{equation}
which does not satisfy the single-OAM-sector condition
$\eta_{\rm OAM}\ll1$. Relative narrowness instead requires
$\eta_{\rm OAM}/|mp|\ll1$. The rms displacement is
$L\sigma_k/k=\SI{15}{nm}$ and the first-order separation for normal
incidence is $2\pi L/(kD)\simeq\SI{9.42}{nm}$. Thus spatial isolation
of neighboring orders is not ensured by this example. In the additional
limiting example $\zeta=0$, $\tilde\chi_{\rm em}=0$, one has
$\rho_{\rm c}\simeq\SI{36.9}{nm}$ and $\eta_{\rm prof}\simeq0.406$.

\section{Discussion}
\label{sec:discussion}

The results separate three different measures of the quality of a generated
electron vortex state. The fork grating fixes the mean OAM of the selected
order, the residual transverse momentum spread determines the statistical
OAM width and transverse purity, and the subsequent propagation determines
how clearly the vortex structure appears in the measured intensity profile.
These quantities therefore characterize different aspects of the same
prepared beam.

From Eq.~\eqref{eq:dispersion_final}, the absolute rms OAM width is
\begin{equation}
\eta_{\rm OAM}\equiv\frac{\Delta L_{z,m}}{\hbar}
=\sigma_k\sqrt{\frac{\zeta^2+z_R^2}{2kz_R}}.
\label{eq:eta_oam}
\end{equation}
For a nonzero-charge order, the relative dispersion is
\begin{equation}
\eta_{\rm rel,m}\equiv
\frac{\Delta L_{z,m}}{|\langle L_z\rangle_m|}
=\frac{\eta_{\rm OAM}}{|mp|},
\qquad mp\ne0.
\label{eq:eta_relative}
\end{equation}
A relatively narrow OAM distribution therefore requires
$\eta_{\rm rel,m}\ll1$. The stronger condition
$\eta_{\rm OAM}\ll1$ corresponds to a sub-$\hbar$ absolute width around the
nominal OAM eigenvalue. For $|mp|=1$, the two criteria coincide.

The distinction becomes especially important for large $|mp|$. A state may
satisfy
\begin{equation}
\hbar\ll\Delta L_{z,m}\ll |mp|\hbar,
\label{eq:semiclassical_oam_window}
\end{equation}
so that its OAM fluctuations extend over many quantum units while remaining
small compared with the mean value. In this regime the relative OAM
distribution is narrow, whereas the absolute fluctuations have a
semiclassical character. The value of $\langle L_z\rangle_m$ alone therefore
does not determine how sharply the transverse quantum state is prepared.

Related optical descriptions treat spatial coherence and OAM jointly
through coherence--OAM matrices \cite{korotkova2021}. Analytical studies of
partially coherent optical vortex beams also show that the correlation
function affects the OAM spectrum \cite{wang2026}.

For the Gaussian electron ensemble considered here, statistical mixing is
quantified by the transverse purity. For the normalized density operator
\begin{equation}
\hat\rho_m=\int d^2k_\perp\,P(\bm k_\perp)
|\psi_{m,\bm k_\perp}\rangle
\langle\psi_{m,\bm k_\perp}|,
\label{eq:density_matrix}
\end{equation}
where $|\psi_{m,\bm k_\perp}\rangle$ is a normalized component state.
At the grating, the components share the Gaussian intensity
$|\psi_{m,\bm k_\perp}|^2\propto e^{-2r^2/w_g^2}$ and differ by their
linear tilt phases. The common radial and vortex phases cancel in the
overlap. The Fourier integral of the normalized Gaussian intensity gives
\begin{equation}
\left|\langle\psi_{m,\bm k_\perp}|\psi_{m,\bm k'_\perp}\rangle\right|^2
=\exp\left[-\frac{w_g^2}{4}|\bm k_\perp-\bm k'_\perp|^2\right].
\label{eq:component_overlap}
\end{equation}
Common unitary Fresnel propagation preserves this overlap. Evaluating
$\operatorname{Tr}(\hat\rho_m^2)$ averages
Eq.~\eqref{eq:component_overlap} over two independent draws from
Eq.~\eqref{eq:momentum_distribution}. Their difference
$\bm\delta=\bm k_\perp-\bm k'_\perp$ has density
$e^{-\delta^2/(2\sigma_k^2)}/(2\pi\sigma_k^2)$, where
$\delta=|\bm\delta|$. Hence
\begin{align}
\mathcal P_{\perp,m}
&=\int\frac{d^2\bm\delta}{2\pi\sigma_k^2}
\exp\left[-\frac{(2+\sigma_k^2w_g^2)\delta^2}{4\sigma_k^2}\right]
\notag\\
&=\frac{1}{1+\sigma_k^2w_g^2/2}.
\label{eq:purity_average}
\end{align}
Using $w_g^2=2(\zeta^2+z_R^2)/(kz_R)$ and
Eq.~\eqref{eq:dispersion_final} gives
\begin{align}
\mathcal P_{\perp,m}
&\equiv\operatorname{Tr}(\hat\rho_m^{\,2})
=\left[1+\frac{\sigma_k^2(\zeta^2+z_R^2)}{kz_R}\right]^{-1}
\notag\\
&=\left[1+2\frac{\Delta L_{z,m}^2}{\hbar^2}\right]^{-1}
=\frac{1}{1+2\eta_{\rm OAM}^2}
\notag\\
&=\frac{1}{1+2(mp)^2\eta_{\rm rel,m}^2},
\qquad mp\ne0.
\label{eq:purity}
\end{align}
The absolute-dispersion form also applies to $mp=0$. Purity approaches unity
for a nearly pure transverse state. The relation in
Eq.~\eqref{eq:purity} is specific to the Gaussian tilt mixture and the
full-plane order model.

Equation~\eqref{eq:purity} also shows that increasing $|mp|$ at fixed
illumination decreases the relative OAM dispersion while leaving the purity
unchanged. A beam with a large mean OAM can therefore have a small fractional
OAM width and substantial statistical mixing at the same time. This
distinction is relevant for interference measurements, since a visually
well-defined annular profile or a small value of
$\Delta L_{z,m}/|\langle L_z\rangle_m|$ does not by itself imply high
transverse coherence.

The propagated intensity profile provides an additional and independent
criterion. The rms spread of the component centers on the observation plane
is $(z-\zeta)\sigma_k/k$. A useful profile-blurring parameter is therefore
\begin{equation}
\eta_{\rm prof}\equiv
\frac{(z-\zeta)\sigma_k}{k\rho_{\rm c}(z)}.
\label{eq:eta_profile}
\end{equation}
The condition $\eta_{\rm prof}\ll1$ indicates weak blurring when
$\rho_{\rm c}$ represents the visible ring scale. Since $\rho_{\rm c}$ is an
asymptotic exponential scale rather than an exact ring radius, quantitative
visibility should be evaluated from the complete radial profile,
particularly near the grating and for $|mp|\ne1$.

The different propagation dependence of
Eqs.~\eqref{eq:eta_oam} and \eqref{eq:eta_profile} is physically important.
The full-plane OAM variance is fixed by the state prepared at the grating and
remains unchanged during subsequent free propagation. The measured annular
profile evolves with the observation distance because the centers of the
different transverse-momentum components separate. A filled central minimum
can therefore coexist with unchanged OAM moments.

The momentum width $\sigma_k$ entering these expressions should be
interpreted as the residual angular width of the phase-space fraction that
reaches the fork grating. It need not coincide with the width inferred from
the MTE of the full photocathode distribution. Angular selection and
apertures reduce $\sigma_k$ at the cost of transmitted current and thereby
increase both OAM concentration and transverse purity
\cite{vanOudheusden2010coherence,vanMourik2014}. The formulas above should
therefore be viewed as constraints on the beam prepared at the mask.

The deterministic grating momentum has a different role. It determines the
displacement and spatial separation of the diffraction orders. After the
coordinate translation and momentum subtraction defining the common order
frame, it does not contribute to $\eta_{\rm OAM}$. The grating period $D$ and
the propagation distance $z-\zeta$ can consequently be adjusted to improve
order separation and screen visibility without changing the intrinsic
full-plane OAM variance, provided that the selected order remains
untruncated.

The electromagnetic Hartree correction affects the radial propagation
through $\tilde\gamma$ and $\rho_{\rm c}$ in
Eqs.~\eqref{eq:gamma_q} and \eqref{eq:rho_cross_general}. Its positive
quadratic curvature produces an outward transverse wave-vector increment
and therefore a defocusing correction. The magnetic self-field reduces the
electric contribution through the factor $1/\gamma^2$. The common
rotationally symmetric phase preserves the centered OAM moments and
transverse purity.

Purity conservation also holds beyond the phase-screen approximation:
evolution under a common Hermitian Hartree Hamiltonian preserves the
spectrum of the density operator and hence
$\mathcal P_\perp=\operatorname{Tr}(\hat\rho_\perp^2)$, including when the
field is updated self-consistently. Axial symmetry additionally preserves
the full-plane OAM moments during this evolution. Stronger self-field
evolution before the grating can change the beam radius and
position-momentum correlations, so the incident state must then be obtained
from a self-consistent propagation calculation. Equation~\eqref{eq:purity}
applies when that state has the assumed Gaussian tilt-mixture form. A change
in the purity of a reduced or transmitted state requires additional
physics, such as coupling to unresolved degrees of freedom, correlations
beyond the Hartree approximation, or state-dependent losses and
postselection.

The present model applies to a quasistatic, paraxial, weak-self-field regime
in which the interaction acts primarily as a quadratic phase screen.
Exchange and correlation corrections, nonlinear Hartree evolution,
longitudinal structure, energy spread, upstream position-momentum
correlations, radiation, aberrations, and finite-aperture truncation lie
outside this approximation. Their inclusion requires self-consistent
source-to-grating propagation.

Weak departures from cylindrical symmetry provide a natural extension.
An elliptic beam density produces an anisotropic Hartree field whose
quadratic part acts as an astigmatic phase. Such a phase can redistribute the
state among OAM modes and generate mode conversion while preserving the
total transverse purity when it is common to the ensemble. Astigmatic mode
conversion of electron vortices has been demonstrated using electron-optical
elements \cite{schattschneider2012}. The evolution of twisted states under
broken axial symmetry has been studied in magnetic
fields \cite{filina2024}, suggesting a related mechanism driven by
anisotropic self-fields.

Within the regime considered here, the analytical separation between mean
OAM, absolute OAM width, transverse purity, and profile visibility provides
a compact set of design criteria for fork-grating preparation of electron
vortex states.

\section{Conclusion}

We have derived an analytical description of fork-grating diffraction for a
Gaussian electron ensemble with finite transverse momentum spread and a
first electromagnetic Hartree phase accumulated before the grating. In the
common order frame, the mean OAM remains fixed at
$\langle L_z\rangle_m=mp\hbar$, while the variance is determined by the
residual transverse momentum spread and the illumination radius at the
grating. For the Gaussian mixture, the transverse purity is directly related
to this variance through
\begin{equation}
\mathcal P_{\perp,m}
=
\left[1+2\left(\frac{\Delta L_{z,m}}{\hbar}\right)^2\right]^{-1}.
\end{equation}

The same transverse momentum spread also produces a propagation-dependent
blurring of the observed annular profile. The OAM variance, transverse
purity, and screen visibility therefore characterize distinct properties of
the generated state. Angular selection improves OAM concentration and
purity at the cost of transmitted current, while the grating period and
post-grating propagation distance control order separation and profile
visibility.

Within the radial phase-screen approximation, the electromagnetic self-field
modifies the propagated radial profile and preserves the full-plane OAM
moments and transverse purity. A self-consistent treatment becomes necessary
when space-charge evolution significantly changes the beam radius or
transverse momentum distribution before diffraction. This provides the
natural next step toward quantitative modeling of high-current vortex-beam
generation.

At large $|mp|$, the regime
$\hbar\ll\Delta L_{z,m}\ll|mp|\hbar$ in
Eq.~\eqref{eq:semiclassical_oam_window} combines a small relative OAM
dispersion with low transverse purity [Eq.~\eqref{eq:purity}]. Such a beam
has semiclassical OAM statistics and can retain an annular intensity profile
with substantial statistical mixing of its transverse state. Applications
that rely on coherent interference therefore require an assessment of
transverse coherence in addition to the mean OAM, relative dispersion, and
visible ring structure.

\appendix

\section{Relativistic reduction for the beam self-field}
\label{app:relativistic_reduction}

Here we give the complete sequence of approximations leading from the Dirac equation to Eq.~\eqref{eq:paraxial_sp}. Up to the point at which terms are explicitly discarded, no expansion in powers of $p/(m_ec)$ is made.

For a static mean field, use the time factor $e^{+iEt/\hbar}$ and
the matrix convention specified in Sec.~\ref{sec:model}. In particular,
$\sigma_i\sigma_j=\delta_{ij}\mathbb I-i\epsilon_{ijk}\sigma_k$.
The upper and lower components of the stationary Dirac spinor obey
\begin{align}
(E+e\Phi_H-m_ec^2)\varphi
&=c\bm\sigma\cdot\hat{\bm\pi}_H\chi,
\notag\\
(E+e\Phi_H+m_ec^2)\chi
&=c\bm\sigma\cdot\hat{\bm\pi}_H\varphi.
\label{eq:coupled_dirac_app}
\end{align}
With $\mathcal G=E+e\Phi_H+m_ec^2$, the lower component is exactly
\begin{equation}
\chi=\frac{c}{\mathcal G}
\bm\sigma\cdot\hat{\bm\pi}_H\varphi.
\label{eq:lower_component_app}
\end{equation}
The position-dependent factor $\mathcal G^{-1}$ does not commute with $\hat{\bm\pi}_H$. Using
\begin{equation}
(\bm\sigma\cdot\hat{\bm\pi}_H)^2
=\hat{\bm\pi}_H^2+e\hbar\bm\sigma\cdot\bm B_H
\label{eq:pauli_identity_app}
\end{equation}
and
\begin{equation}
\mathcal G\left[\bm\sigma\cdot\hat{\bm\pi}_H,
\mathcal G^{-1}\right]
=-\frac{ie\hbar}{\mathcal G}\bm\sigma\cdot\nabla\Phi_H,
\label{eq:G_commutator_app}
\end{equation}
one obtains the exact upper-component equation
\begin{align}
\Bigg[\hat{\bm\pi}_H^2+e\hbar\bm\sigma\cdot\bm B_H
&+\frac{e\hbar}{\mathcal G}
\Big\{-i\nabla\Phi_H\cdot\hat{\bm\pi}_H
\notag\\
&-\bm\sigma\cdot
(\nabla\Phi_H\times\hat{\bm\pi}_H)\Big\}\Bigg]\varphi
\notag\\
&=\frac{(E+e\Phi_H)^2-m_e^2c^4}{c^2}\varphi.
\label{eq:quadratic_dirac_explicit_app}
\end{align}
The three correction operators are explicitly
\begin{align}
\hat{\mathcal Z}_H&=e\hbar\bm\sigma\cdot\bm B_H,
\notag\\
\hat{\mathcal S}_H&=-\frac{e\hbar}{\mathcal G}
\bm\sigma\cdot(\nabla\Phi_H\times\hat{\bm\pi}_H),
\notag\\
\hat{\mathcal D}_H&=-\frac{ie\hbar}{\mathcal G}
\nabla\Phi_H\cdot\hat{\bm\pi}_H.
\label{eq:correction_operators_app}
\end{align}
The last operator is spin independent. In a low-energy Foldy-Wouthuysen expansion, its Hermitian reorganization contains the Darwin contribution, whereas $\hat{\mathcal S}_H$ produces the usual electric spin-orbit term. Equation~\eqref{eq:quadratic_dirac_explicit_app}, however, is not restricted to that expansion.

For the cylindrical beam, $\nabla\Phi_H$ is transverse, while the dominant mechanical momentum is longitudinal. Consequently, the spin-orbit operator is controlled by $|\nabla\Phi_H|p_z$, whereas the scalar-gradient operator contains the smaller product $|\nabla\Phi_H|p_\perp$. Writing
\begin{align}
|\nabla\Phi_H|&\sim\frac{|\Phi_H|}{w_0},
&
|\bm B_H|&\sim\frac{v_z}{c^2}\frac{|\Phi_H|}{w_0},
\notag\\
\mathcal G&\simeq(\gamma+1)m_ec^2,
\label{eq:self_field_scales_app}
\end{align}
and comparing with the scalar linear term $2Ee\Phi_H/c^2$ gives
\begin{align}
\frac{|\hat{\mathcal Z}_H|}{|2Ee\Phi_H/c^2|}
&\sim\frac{\lambda_C}{2\gamma w_0}\frac{v_z}{c},
\notag\\
\frac{|\hat{\mathcal S}_H|}{|2Ee\Phi_H/c^2|}
&\sim\frac{\lambda_C}{2(\gamma+1)w_0}\frac{v_z}{c},
\notag\\
\frac{|\hat{\mathcal D}_H|}{|2Ee\Phi_H/c^2|}
&\sim\frac{\lambda_C}{2\gamma(\gamma+1)w_0}
\frac{p_\perp}{m_ec}.
\label{eq:operator_estimates_app}
\end{align}
These are the three individual estimates summarized in the main text.

Here $\lambda_C=\hbar/(m_ec)$ is the reduced Compton wavelength.
For a tilted component, $p_\perp$ includes the mechanical tilt momentum
as well as the intrinsic Gaussian width. At high $\gamma$, comparison
with the net orbital self-field must also account for electric-magnetic
cancellation; Eq.~\eqref{eq:operator_estimates_app} alone is not such a
comparison.

For a Gaussian component with $p_\perp\sim\hbar/w_0$, the final ratio is of order $(\lambda_C/w_0)^2$. Dropping Eqs.~\eqref{eq:correction_operators_app} under these conditions leaves the scalar relativistic Helmholtz equation
\begin{equation}
\hat{\bm\pi}_H^2\varphi
=\frac{(E+e\Phi_H)^2-m_e^2c^4}{c^2}\varphi.
\label{eq:scalar_relativistic_helmholtz}
\end{equation}

We now make the longitudinal-envelope approximation. For the geometry in Eq.~\eqref{eq:self_field_geometry}, assume $\partial_zA_{H,z}=0$ over the interaction region and substitute $\varphi_s=u(\bm r_\perp,z)e^{-ikz}\eta_s$ into Eq.~\eqref{eq:scalar_relativistic_helmholtz}. After using $E^2=m_e^2c^4+\hbar^2c^2k^2$, but before either the weak-field or paraxial truncation, the exact envelope equation is
\begin{align}
-\hbar^2\nabla_\perp^2u
&+2i\hbar^2k\partial_zu
-\hbar^2\partial_z^2u
+2e\hbar kA_{H,z}u
\notag\\
&+2ie\hbar A_{H,z}\partial_zu
+e^2A_{H,z}^2u
\notag\\
&=\frac{2Ee}{c^2}\Phi_Hu
+\frac{e^2}{c^2}\Phi_H^2u.
\label{eq:full_envelope_app}
\end{align}
This equation makes the remaining omissions explicit. The weak-field conditions $e|\Phi_H|/E\ll1$ and $e|A_{H,z}|/p_z\ll1$ remove $\Phi_H^2$ and $A_{H,z}^2$. The paraxial condition $|\partial_z^2u|\ll|2k\partial_zu|$ removes the second longitudinal derivative. Finally, $A_{H,z}\partial_zu$ is the product of a weak-field factor and the small longitudinal-envelope variation, and is beyond the retained combined order. The resulting intermediate equation is
\begin{equation}
-\hbar^2\nabla_\perp^2u+2i\hbar^2k\partial_zu
+2e\hbar kA_{H,z}u
=\frac{2Ee}{c^2}\Phi_Hu.
\label{eq:paraxial_intermediate_app}
\end{equation}
Solving for $\partial_zu$ and using $v_z=\hbar kc^2/E$ yields
\begin{equation}
\partial_zu=-\frac{i}{2k}\nabla_\perp^2u
-\frac{ie}{\hbar v_z}\Phi_Hu+\frac{ie}{\hbar}A_{H,z}u,
\label{eq:paraxial_explicit_app}
\end{equation}
which is Eq.~\eqref{eq:paraxial_sp}.

There is no nonrelativistic approximation hidden in this last step. The longitudinal wave number $k$ is related to the total energy by the exact relativistic dispersion. If that dispersion were expanded at low momentum, it would generate the Foldy-Wouthuysen free correction $-\hat{\bm p}^{\,4}/(8m_e^3c^2)$. Here it is already included through $E(k)$. By contrast, solving the free Helmholtz equation as an expansion in the transverse angle gives
\begin{equation}
\partial_zu=-\frac{i}{2k}\nabla_\perp^2u
+\frac{i}{8k^3}\nabla_\perp^4u
+O\!\left(\frac{\nabla_\perp^6}{k^5}\right),
\label{eq:nonparaxial_correction_app}
\end{equation}
so the $\nabla_\perp^4$ term is omitted because it is nonparaxial, not because it is relativistic.

For a uniform cylindrical beam, the transverse Maxwell equations are
\begin{equation}
\frac{1}{r}\partial_r(r\partial_r\Phi_H)=\frac{en_0}{\varepsilon_0},
\qquad
\frac{1}{r}\partial_r(r\partial_rA_{H,z})=\mu_0en_0v_z.
\label{eq:cylindrical_maxwell_app}
\end{equation}
Their regular interior solutions are Eq.~\eqref{eq:uniform_hartree_potentials}. The phase-driving combination is then
\begin{align}
\Phi_H-v_zA_{H,z}
&=\text{const}+\frac{en_0r^2}{4\varepsilon_0}
\left(1-\frac{v_z^2}{c^2}\right)
\notag\\
&=\text{const}+\frac{en_0r^2}{4\varepsilon_0\gamma^2}.
\label{eq:self_field_cancellation_app}
\end{align}
Thus the magnetic self-field does not introduce a new transverse
functional form for a uniform beam. It changes the curvature of the
quadratic phase accumulated before the grating, while the subsequent
free Fresnel diffraction calculation remains unchanged.

The last approximation used in the main text is the first-Hartree eikonal phase screen in Eq.~\eqref{eq:eikonal_hartree_main}. The potentials are evaluated from the noninteracting density and current and are not updated after the wave function acquires the self-field phase. Free diffraction remains included in $u_{\bm k_\perp}^{(0)}$, but the action of the transverse Laplacian on the additional phase, and hence the self-field-induced change of the transverse amplitude, is neglected. This step is valid when the mean field changes the beam width only weakly over the interval $0<z<\zeta$.

This is an extra phase-screen approximation to frozen-field propagation,
not its exact first-order solution. A useful sufficient displacement
estimate for a nearly constant quadratic field is
$2\tilde\chi_{\rm em}\zeta/k\ll1$; beam diffraction and tilt walkoff must
also be consistent with the prepared boundary condition. Without those
conditions, the accumulated phase must be evaluated along the evolving
trajectories or by solving the paraxial equation with the frozen field.

\section{Fresnel evaluation of the diffracted order}
\label{app:fresnel}

After the grating the field is written in Eq.~\eqref{eq:fresnel_app}
\begin{align}
u(\rho,\theta,z)=&\,\frac{ik e^{-ik(z-\zeta)}}{2\pi(z-\zeta)}
\int d^2r\,u_{\rm in}(r,\varphi,\zeta)T(r,\varphi)
\notag\\
&\times\exp\left[-\frac{ik}{2(z-\zeta)}|\bm r_\perp-\bm\rho|^2\right].
\end{align}
To display the steps leading to Eq.~\eqref{eq:um_final}, write
$L=z-\zeta$ and $\ell=mp$. Inserting
Eqs.~\eqref{eq:incident_field} and \eqref{eq:fork_transmission} into
Eq.~\eqref{eq:fresnel_app}, collecting the terms quadratic in the
grating-plane radius, and using Eq.~\eqref{eq:gamma_q} gives the radial
factor
\begin{equation}
\exp\left[-\frac{ik\tilde\gamma r^2}{2}\right].
\label{eq:quadratic_factor_app}
\end{equation}
The remaining terms linear in $r\cos\varphi$ and $r\sin\varphi$ are
combined by the shifted coordinates in
Eqs.~\eqref{eq:order_shift} and \eqref{eq:shifted_coords}. With these
definitions, the $m$th-order contribution can be written as
\begin{align}
u_m={}&\,\mathcal A_m(z)
\exp\left[-ik\left(z+\frac{\rho^2}{2L}\right)\right]
\notag\\
&\times\int_0^\infty r\,dr\,
\exp\left[-\frac{ik\tilde\gamma r^2}{2}\right]
\notag\\
&\times\int_0^{2\pi}d\varphi\,
\exp\left[\frac{ik\rho_mr}{L}
\cos(\varphi-\theta_m)\right]e^{-i\ell\varphi},
\label{eq:order_integral_app}
\end{align}
where
\begin{equation}
\mathcal A_m(z)=\frac{ik}{2\pi L}\,
t_m\frac{q(0)}{q(\zeta)}.
\label{eq:order_prefactor_app}
\end{equation}
Thus the grating momentum and the incident transverse momentum enter
the integral only through the displaced polar pair
$(\rho_m,\theta_m)$, whereas the Gaussian and Hartree curvatures enter
through $\tilde\gamma$.

The angular integration follows from the Jacobi-Anger expansion
\begin{equation}
e^{ia\cos\chi}=\sum_{s=-\infty}^{\infty}
i^sJ_s(a)e^{is\chi},
\label{eq:jacobi_anger_app}
\end{equation}
which gives eq. ~\eqref{eq:angular_integral_app}
\begin{equation}
\int_0^{2\pi}e^{ia r\cos(\varphi-\theta_m)}e^{-i\ell\varphi}\,d\varphi
=2\pi i^{\lvert\ell\rvert}J_{\lvert\ell\rvert}(ar)e^{-i\ell\theta_m}
\end{equation}
for integer $\ell$; here
$i^\ell J_\ell(ar)=i^{\lvert\ell\rvert}J_{\lvert\ell\rvert}(ar)$
also covers negative diffraction orders. Consequently, all remaining
transverse structure is
contained in the radial integral as in eq. \eqref{eq:radial_integral_app}
\begin{equation}
Y_m(\rho_m)=\int_0^\infty J_{\lvert\ell\rvert}\left(\frac{k\rho_m r}{z-\zeta}\right)
\exp\left[-\frac{ik\tilde\gamma r^2}{2}\right]r\,dr.
\end{equation}

As in the Gaussian fork-grating calculation of
Ref.~\cite{janicijevic2008}, this integral may first be reduced to a
confluent hypergeometric function. Set
\begin{equation}
A=\frac{ik\tilde\gamma}{2},
\qquad
B=\frac{k\rho_m}{L}.
\label{eq:AB_radial_app}
\end{equation}
For $\operatorname{Re}A>0$, the standard Bessel-Gaussian integral is
\begin{align}
&\int_0^\infty r e^{-Ar^2}J_{\lvert\ell\rvert}(Br)\,dr
={}\frac{B^{\lvert\ell\rvert}}
{2^{\lvert\ell\rvert+1}A^{\lvert\ell\rvert/2+1}}
\frac{\Gamma(\lvert\ell\rvert/2+1)}
{\Gamma(\lvert\ell\rvert+1)}
\notag\\
&\times M\left(\frac{\lvert\ell\rvert}{2}+1;
\lvert\ell\rvert+1;
-\frac{B^2}{4A}\right),
\label{eq:radial_integral_kummer_app}
\end{align}
where $M(a;b;x)={}_1F_1(a;b;x)$. Applying the Kummer transformation
\begin{equation}
M(a;b;-s)=e^{-s}M(b-a;b;s)
\label{eq:kummer_transformation_app}
\end{equation}
and then the identity
\begin{align}
&M\left(\frac{\lvert\ell\rvert}{2};
\lvert\ell\rvert+1;s\right)
={}\frac{\sqrt{\pi}\,\Gamma(\lvert\ell\rvert+1)}
{2\Gamma(\lvert\ell\rvert/2+1)}
s^{(1-\lvert\ell\rvert)/2}e^{s/2}
\notag\\
&\times\left[
I_{(\lvert\ell\rvert-1)/2}\left(\frac{s}{2}\right)
-I_{(\lvert\ell\rvert+1)/2}\left(\frac{s}{2}\right)
\right]
\label{eq:kummer_bessel_app}
\end{align}
gives the equivalent Weber-Sonine form
\begin{align}
&\int_0^\infty r e^{-Ar^2}J_{\lvert\ell\rvert}(Br)\,dr
={}\frac{\sqrt{\pi}B}{8A^{3/2}}
\exp\left(-\frac{B^2}{8A}\right)
\notag\\
&\times\left[
I_{(\lvert\ell\rvert-1)/2}\!\left(\frac{B^2}{8A}\right)
-I_{(\lvert\ell\rvert+1)/2}\!\left(\frac{B^2}{8A}\right)
\right].
\label{eq:weber_sonine_app}
\end{align}

For the parameters in Eq.~\eqref{eq:AB_radial_app}, define
\begin{equation}
\xi_m=\frac{k\rho_m^2}{8Q(z)}.
\label{eq:xi_radial_app}
\end{equation}
Because
\begin{equation}
\frac{B^2}{8A}=-i\xi_m,
\qquad
I_\nu(-i\xi_m)=i^{-\nu}J_\nu(\xi_m),
\label{eq:I_to_J_app}
\end{equation}
Eq.~\eqref{eq:weber_sonine_app} becomes
\begin{align}
&Y_m(\rho_m)
={}\frac{\sqrt{\pi}k\rho_m}
{8(z-\zeta)(ik\tilde\gamma/2)^{3/2}}
i^{-(\lvert\ell\rvert-1)/2}\exp\left[\frac{ik\rho_m^2}{8Q(z)}\right]
\notag\\
&\times
\Bigg[
J_{(\lvert\ell\rvert-1)/2}\!\left(
\frac{k\rho_m^2}{8Q(z)}\right)
+iJ_{(\lvert\ell\rvert+1)/2}\!\left(
\frac{k\rho_m^2}{8Q(z)}\right)
\Bigg].
\label{eq:radial_integral_bessel_app}
\end{align}
The relative factor $+i$ between the two ordinary Bessel functions
comes from the one-unit difference between their orders in
Eq.~\eqref{eq:I_to_J_app}. Combining
Eq.~\eqref{eq:radial_integral_bessel_app} with the prefactor in
Eq.~\eqref{eq:order_prefactor_app} and the angular factor in
Eq.~\eqref{eq:angular_integral_app} yields Eq.~\eqref{eq:um_final}.

\section{Asymptotic envelope of the first vortex order}
\label{app:asymptotic_profile}

For $m=p=1$, set $\Xi=k\rho_m^2/[8Q(z)]$. The radial part is
\begin{equation}
u_1\propto \rho_m e^{i\Xi}\left[J_0(\Xi)+iJ_1(\Xi)\right].
\label{eq:u1_radial_app}
\end{equation}
Using $J_\nu=(H_\nu^{(1)}+H_\nu^{(2)})/2$ and
\begin{align}
H_\nu^{(1)}(\Xi)&\sim\sqrt{\frac{2}{\pi\Xi}}e^{i(\Xi-\nu\pi/2-\pi/4)},
\notag\\
H_\nu^{(2)}(\Xi)&\sim\sqrt{\frac{2}{\pi\Xi}}e^{-i(\Xi-\nu\pi/2-\pi/4)},
\label{eq:hankel_asym_app}
\end{align}
one finds that the leading $e^{-i\Xi}$ contribution cancels in
$J_0+iJ_1$, while the $e^{+i\Xi}$ contribution remains. Including
the preceding exponential, this branch contributes the intensity
factor $e^{-4\operatorname{Im}\Xi}$, where
\begin{equation}
-4\operatorname{Im}\Xi
=-\frac{k\rho_m^2z_R}
{\left|q(z)+2\tilde\chi_{\rm em}(z-\zeta)q(\zeta)/k\right|^2}
=-\frac{\rho_m^2}{4\rho_{\rm c}^2}.
\label{eq:u1_raw_envelope_app}
\end{equation}
This gives the scale in Eq.~\eqref{eq:rho_cross}. The subleading
branch produces an algebraic tail; $\rho_{\rm c}$ characterizes the
exponential contribution, not the exact ring maximum. The large-argument
expansion cannot be extrapolated to the near-axis limit.

For nonzero charge the ideal phase-only mask also gives an algebraic
far tail. For the first order, $Y_1\sim B^{-2}$ and the intensity behaves
as $\rho_m^{-4}$ at fixed propagation distance.

\section{Centered OAM operator}
\label{app:oam_operator}

For calculations around the center of the selected diffraction order it is useful to set the reference-order shift to zero and keep only the residual displacement of an individual transverse-momentum component. Writing
\begin{equation}
\rho_m\cos\theta_m+\alpha_m=\rho\cos\theta,
\qquad
\rho_m\sin\theta_m+\beta_m=\rho\sin\theta,
\label{eq:centered_coords_app}
\end{equation}
one obtains
\begin{equation}
\frac{-i}{\hbar}\hat L_z\equiv D_m=N_-\partial_{\rho_m}
+\left(1+\frac{N_+}{\rho_m}\right)\partial_{\theta_m},
\label{eq:Dmapp}
\end{equation}
where
\begin{align}
N_+&=\alpha_m\cos\theta_m+\beta_m\sin\theta_m,
\notag\\
N_-&=\alpha_m\sin\theta_m-\beta_m\cos\theta_m.
\label{eq:centered_N_app}
\end{align}
The nonconstant coefficients give
\begin{align}
D^2=&\left(N_+ +\frac{N_+^2}{\rho_m}\right)\partial_{\rho_m}
+\left(-\frac{N_-}{\rho_m}-\frac{2N_-N_+}{\rho_m^2}\right)\partial_{\theta_m}
\notag\\
&+\left(2N_-+\frac{2N_-N_+}{\rho_m}\right)\partial_{\rho_m}\partial_{\theta_m}
+N_-^2\partial_{\rho_m}^2
\notag\\
&+\left(1+\frac{2N_+}{\rho_m}+\frac{N_+^2}{\rho_m^2}\right)\partial_{\theta_m}^2.
\label{eq:D2_app}
\end{align}
This is the operator used in Eq.~\eqref{eq:lz2_def}, with $\hat L_z^2=-\hbar^2D^2$.
With $\ell=mp$, the radial factor read directly from
Eq.~\eqref{eq:um_final} is
\begin{align}
\label{eq:centered_short_notation}
&R_\ell(\rho_m,z)
=\rho_m
\exp\left[\frac{ik\rho_m^2}{8Q(z)}\right]
F\!\left(\frac{k\rho_m^2}{8Q(z)}\right)=
\\
&\rho_m
\exp\left[\frac{ik\rho_m^2}{8Q(z)}\right]
\left[
J_{(|\ell|-1)/2}\!\left(\frac{k\rho_m^2}{8Q(z)}\right)
+iJ_{(|\ell|+1)/2}\!\left(\frac{k\rho_m^2}{8Q(z)}\right)
\right] \nonumber.
\end{align}
Up to a $\rho_m$-independent normalization, the centered component is
\begin{align}
u_m={}&\mathcal C(z)
\exp\left[-\frac{ik}{2(z-\zeta)}
\left(\rho_m^2+\alpha_m^2+\beta_m^2
+2\rho_mN_+\right)\right]
\notag\\
&\times e^{-i\ell\theta_m}R_\ell(\rho_m,z).
\label{eq:centered_radial_factorization}
\end{align}
Substitution into Eq.~\eqref{eq:Dmapp} gives directly
\begin{align}
Du_m
={}&\mathcal C
\exp\left[-\frac{ik}{2(z-\zeta)}
\left(\rho_m^2+\alpha_m^2+\beta_m^2
+2\rho_mN_+\right)\right]
\notag\\
&\times e^{-i\ell\theta_m}
\left[
N_-R_\ell'
-i\ell\left(1+\frac{N_+}{\rho_m}\right)R_\ell
\right].
\label{eq:D_simplified_app}
\end{align}
All terms proportional to $k/(z-\zeta)$ cancel before the square is taken.
They belong to one common translated Fresnel phase and cannot be
interpreted as independent positive contributions to the dispersion.
Using
\begin{align}
\int_0^{2\pi}N_\pm\,d\theta_m&=0,
&
\int_0^{2\pi}N_-N_+\,d\theta_m&=0,
\notag\\
\int_0^{2\pi}N_\pm^2\,d\theta_m
&=\pi(\alpha_m^2+\beta_m^2).
\label{eq:N_angular_integrals_app}
\end{align}
we obtain the first final form,
\begin{align}
&\int_0^{2\pi}\left(
|Du_m|^2-\ell^2|u_m|^2
\right)d\theta_m
\notag\\
&\qquad=|\mathcal C|^2\pi(\alpha_m^2+\beta_m^2)\left(
|R_\ell'|^2
+\frac{\ell^2}{\rho_m^2}|R_\ell|^2
\right).
\label{eq:Du_angular_average_app}
\end{align}
Here $\langle L_z\rangle_m(\bm k_\perp)=\ell\hbar$, so the left-hand
side of Eq.~\eqref{eq:Du_angular_average_app} is specifically the
angularly integrated numerator of the OAM variance
$\Delta L_{z,m}^2=\langle L_z^2\rangle_m-\langle L_z\rangle_m^2$,
divided by $\hbar^2$. It is not merely the second moment. The
right-hand side is manifestly real and positive. Radial integration
therefore gives the component dispersion directly:
\begin{equation}
\frac{\Delta L_{z,m}^2(\bm k_\perp)}{\hbar^2}
=\frac{\alpha_m^2+\beta_m^2}{2}
\frac{\displaystyle\int_0^\infty \rho_m\,d\rho_m
\left[
|R_\ell'|^2+\frac{\ell^2}{\rho_m^2}|R_\ell|^2
\right]}
{\displaystyle\int_0^\infty \rho_m\,d\rho_m\,|R_\ell|^2}.
\label{eq:dispersion_component_radial_app}
\end{equation}
We now evaluate the radial quotient at an arbitrary plane $z>\zeta$,
without assuming propagation invariance. Equations
\eqref{eq:radial_integral_bessel_app} and
\eqref{eq:centered_short_notation} give
\begin{align}
Y_m(\rho_m)
={}&\frac{\sqrt{\pi}k}
{8(z-\zeta)(ik\tilde\gamma/2)^{3/2}}
i^{-(\lvert\ell\rvert-1)/2}R_\ell(\rho_m,z).
\label{eq:Y_R_relation_app}
\end{align}
The prefactor is independent of $\rho_m$ and cancels from the radial
quotient. For an arbitrary radial function $g(r)$, the order-$\nu$
Hankel Parseval identity is
\begin{equation*}
\int_0^\infty\kappa\,d\kappa
\left|\int_0^\infty r\,dr\,g(r)J_\nu(\kappa r)\right|^2
=\int_0^\infty r\,dr\,|g(r)|^2.
\end{equation*}
It follows directly from the closure relation
\begin{equation*}
\int_0^\infty\kappa\,d\kappa\,
J_\nu(\kappa r)J_\nu(\kappa r')
=\frac{\delta(r-r')}{r}.
\end{equation*}
Applying this identity to Eq.~\eqref{eq:radial_integral_app}, with
$\nu=\lvert\ell\rvert$ and
$\kappa=k\rho_m/(z-\zeta)$, gives the first line of
Eq.~\eqref{eq:hankel_integrals_app} below.

For the numerator, the change from $\rho_m$ to $\kappa$ first gives
\begin{align*}
&\int_0^\infty\kappa\,d\kappa\left[
\left|\partial_\kappa
Y_m\!\left(\frac{z-\zeta}{k}\kappa\right)\right|^2
+\frac{\ell^2}{\kappa^2}
\left|Y_m\!\left(\frac{z-\zeta}{k}\kappa\right)\right|^2
\right]
\\
&=\left[
\kappa Y_m^*\!\left(\frac{z-\zeta}{k}\kappa\right)
\partial_\kappa Y_m\!\left(\frac{z-\zeta}{k}\kappa\right)
\right]_0^\infty
\\
&\quad+\int_0^\infty\kappa\,d\kappa\,
Y_m^*\!\left(\frac{z-\zeta}{k}\kappa\right)
\\
&\qquad\times
\left[-\left(
\partial_\kappa^2+\frac{1}{\kappa}\partial_\kappa
-\frac{\ell^2}{\kappa^2}
\right)
Y_m\!\left(\frac{z-\zeta}{k}\kappa\right)
\right].
\end{align*}
The displayed boundary term vanishes because the Hankel transform is
regular at $\kappa=0$ and decays at infinity. The Bessel equation gives
\begin{align*}
&-\left(
\partial_\kappa^2+\frac{1}{\kappa}\partial_\kappa
-\frac{\ell^2}{\kappa^2}
\right)
Y_m\!\left(\frac{z-\zeta}{k}\kappa\right)
\\
&\qquad=\int_0^\infty r^3\,dr\,
e^{-ik\tilde\gamma r^2/2}J_{\lvert\ell\rvert}(\kappa r),
\end{align*}
and a second use of the closure relation reduces the remaining
$\kappa$ integral to the second line of
Eq.~\eqref{eq:hankel_integrals_app}:
\begin{align}
&\int_0^\infty\rho_m\,d\rho_m\,|Y_m|^2
=\frac{(z-\zeta)^2}{k^2}
\int_0^\infty r\,dr\,
\left|e^{-ik\tilde\gamma r^2/2}\right|^2,
\notag\\
&\int_0^\infty\rho_m\,d\rho_m
\left(|Y_m'|^2+\frac{\ell^2}{\rho_m^2}|Y_m|^2\right)
\notag\\
&\qquad=\int_0^\infty r^3\,dr\,
\left|e^{-ik\tilde\gamma r^2/2}\right|^2.
\label{eq:hankel_integrals_app}
\end{align}
Equations~\eqref{eq:Y_R_relation_app} and
\eqref{eq:hankel_integrals_app} therefore give
\begin{align}
&\frac{\displaystyle\int_0^\infty\rho_m\,d\rho_m
\left[|R_\ell'|^2+\dfrac{\ell^2}{\rho_m^2}|R_\ell|^2\right]}
{\displaystyle\int_0^\infty\rho_m\,d\rho_m|R_\ell|^2}
\notag\\
&\qquad=\frac{k^2}{(z-\zeta)^2}
\frac{\displaystyle\int_0^\infty r^3\,dr\,
\left|e^{-ik\tilde\gamma r^2/2}\right|^2}
{\displaystyle\int_0^\infty r\,dr\,
\left|e^{-ik\tilde\gamma r^2/2}\right|^2}.
\label{eq:radial_integrals_arbitrary_z_app}
\end{align}
From Eq.~\eqref{eq:gamma_q},
\begin{align}
&\tilde\gamma
={}\left[\frac{\zeta}{\zeta^2+z_R^2}
+\frac{1}{z-\zeta}+\frac{2\tilde\chi_{\rm em}}{k}\right]
-i\frac{z_R}{\zeta^2+z_R^2},
\notag\\
&\left|e^{-ik\tilde\gamma r^2/2}\right|^2
=e^{-kz_Rr^2/(\zeta^2+z_R^2)}.
\label{eq:hankel_gaussian_weight_app}
\end{align}
The real Hartree and free-curvature terms therefore do not enter the
modulus, and the remaining Gaussian quotient is
\begin{equation}
\frac{\displaystyle\int_0^\infty r^3
e^{-kz_Rr^2/(\zeta^2+z_R^2)}dr}
{\displaystyle\int_0^\infty r
e^{-kz_Rr^2/(\zeta^2+z_R^2)}dr}
=\frac{\zeta^2+z_R^2}{kz_R}.
\label{eq:hankel_gaussian_quotient_app}
\end{equation}
Hence
\begin{equation}
\frac{\displaystyle\int_0^\infty \rho_m\,d\rho_m
\left[|R_\ell'|^2+\dfrac{\ell^2}{\rho_m^2}|R_\ell|^2\right]}
{\displaystyle\int_0^\infty \rho_m\,d\rho_m\,|R_\ell|^2}
=\frac{k^2}{(z-\zeta)^2}
\frac{\zeta^2+z_R^2}{kz_R}.
\label{eq:radial_ratio_arbitrary_z_app}
\end{equation}
Finally,
\begin{equation}
\alpha_m^2+\beta_m^2
=\frac{(z-\zeta)^2}{k^2}(k_x^2+k_y^2),
\label{eq:shift_square_arbitrary_z_app}
\end{equation}
so substitution into Eq.~\eqref{eq:dispersion_component_radial_app}
shows the cancellation at the same observation plane:
\begin{align}
\frac{\Delta L_{z,m}^2(\bm k_\perp)}{\hbar^2}
&=\frac{(z-\zeta)^2(k_x^2+k_y^2)}{2k^2}
\frac{k^2}{(z-\zeta)^2}
\frac{\zeta^2+z_R^2}{kz_R}
\notag\\
&=\frac{k_x^2+k_y^2}{2}
\frac{\zeta^2+z_R^2}{kz_R}.
\label{eq:dispersion_component_identity_app}
\end{align}
Averaging over the transverse-momentum distribution gives
\begin{equation}
\frac{\Delta L_{z,m}^2}{\hbar^2}
=\frac{\sigma_k^2(\zeta^2+z_R^2)}{2kz_R}
=2\sigma_k^2\rho_{\rm c}^2(\zeta).
\label{eq:dispersion_recovered_app}
\end{equation}
This is the exact full-plane result within the stated phase-screen
model, not an estimate.

An independent grating-plane check makes the assumptions transparent.
Write $u_m=g(r)e^{-i\ell\varphi-i\bm k_\perp\cdot\bm r_\perp}$, where
$g$ includes the radial Gaussian curvature and Hartree phase. Then
\begin{align}
\hat L_z u_m
&=\hbar[\ell+x k_y-y k_x]u_m,\notag\\
\langle x^2\rangle=\langle y^2\rangle
&=\frac{w_g^2}{4}
=\frac{\zeta^2+z_R^2}{2kz_R},\notag\\
\frac{\Delta L_z^2(\bm k_\perp)}{\hbar^2}
&=(k_x^2+k_y^2)\frac{w_g^2}{4}.
\label{eq:independent_oam_check}
\end{align}
The radial phase does not enter. Free transverse propagation commutes
with $\hat L_z$, so this result agrees with the full Fresnel/Hankel
calculation at every downstream plane. At $z=\zeta$ the scale
$\rho_{\rm c}^2=(\zeta^2+z_R^2)/(4kz_R)$ is already independent of the
Hartree coefficient; no zero-interaction restriction is needed in
Eq.~\eqref{eq:dispersion_recovered_app}.

\bibliography{refs}

\end{document}